\documentclass[runningheads]{llncs}

\usepackage[utf8]{inputenc}
\usepackage[T1]{fontenc}

\usepackage{float}
\usepackage{makecell}
\usepackage{graphicx} 
\usepackage{booktabs}
\usepackage{blindtext}
\usepackage{url}
\usepackage{textcomp}
\usepackage{hyperref}
\usepackage[misc]{ifsym}

\usepackage[capitalise, nameinlink, noabbrev, sort]{cleveref}
\crefname{lstlisting}{Listing}{Listings}
\Crefname{lstlisting}{Listing}{Listings}

\usepackage{listings}

\usepackage{xcolor}

\definecolor{solarized@base03}{HTML}{002B36}
\definecolor{solarized@base02}{HTML}{073642}
\definecolor{solarized@base01}{HTML}{586e75}
\definecolor{solarized@base00}{HTML}{657b83}
\definecolor{solarized@base0}{HTML}{839496}
\definecolor{solarized@base1}{HTML}{93a1a1}
\definecolor{solarized@base2}{HTML}{EEE8D5}
\definecolor{solarized@base3}{HTML}{FDF6E3}
\definecolor{solarized@yellow}{HTML}{B58900}
\definecolor{solarized@orange}{HTML}{CB4B16}
\definecolor{solarized@red}{HTML}{DC322F}
\definecolor{solarized@magenta}{HTML}{D33682}
\definecolor{solarized@violet}{HTML}{6C71C4}
\definecolor{solarized@blue}{HTML}{268BD2}
\definecolor{solarized@cyan}{HTML}{2AA198}
\definecolor{solarized@green}{HTML}{859900}

\lstnewenvironment{cpp}[1][] {
  \lstset{
  language=c++,
  basicstyle=\ttfamily\scriptsize,
  numbers=left,
  numberstyle=\tiny,
  tabsize=2,
  breaklines=true,
  escapeinside={@}{@},
  numberstyle=\tiny\color{solarized@base01},
  keywordstyle=\color{solarized@green},
  stringstyle=\color{solarized@cyan}\ttfamily,
  commentstyle=\color{solarized@base01},
  emphstyle=\color{solarized@red},
  showstringspaces=false,
  flexiblecolumns=false,
  frame=tb,
  #1
  }
}
{}

\usepackage{microtype}
\usepackage{xspace}

\newcommand{\eg}{e.g.\@\xspace}

\title{Parallel Flowshop in YewPar} \titlerunning{Parallel Flowshop in YewPar}

\author{
Ignas Knizikevi\v{c}ius\inst{1}
\and
Phil Trinder\inst{1}
\and
Blair Archibald$^{\textrm{\Letter}}$\inst{1}
\and
Jinghua Yan\inst{2}
}

\authorrunning{Knizikevi\v{c}ius et al.} 

\institute{
  School of Computing Science, University of Glasgow, Scotland \\
  \email{\{phil.trinder, blair.archibald\}@glasgow.ac.uk} \\
  \and
  School of Computing, University of Utah, United States of America \\
  \email{jhyan@cs.utah.edu}
}

\begin{document}

\maketitle

\begin{abstract}
Parallelism may reduce the time to find exact solutions for many Operations Research (OR) problems, but parallelising combinatorial search is extremely challenging. YewPar is a new combinatorial search framework designed to allow domain specialists to benefit from parallelism by reusing sophisticated parallel search  patterns. 

This paper shows (1) that it is low effort to encode and parallelise a typical OR problem (Flowshop Scheduling FSP) in YewPar even for scalable clusters; (2) that the YewPar library makes it extremely easy to exploit three alternate FSP parallelisations; (3) that the YewPar FSP implementations  are valid, and have sequential performance  comparable  with a published algorithm; and (4) provides a systematic performance evaluation of the three parallel FSP versions on 10 standard FSP instances with up to 240 workers on a Beowulf cluster. 

\end{abstract}

\section{Introduction} \label{sec:Introduction}

Flowshop Scheduling (FSP) is a classic OR problem: \emph{n} jobs are to be processed on \emph{m} machines with the processing time of any job on any machine specified. The goal is to find a sequence of jobs that has the least makespan, or time to complete the jobs. Optimal FSP solutions can be found by exploring the entire search space, but this is computationally expensive. Approximation strategies reduce runtime by exploring only part of the search space and, despite being unable to guarantee optimality, are commonly used.

The runtimes of exact combinatorial search problems like FSP can be reduced (1) by using Branch and Bound (B\&B) techniques that  dynamically explore the search space and discard (prune) branches that cannot contain solutions; and (2) by exploiting parallelism. However parallelising exact combinatorial search is extremely challenging due to the huge and highly irregular search trees, the need to preserve search order heuristics, and pruning that dynamically alters the workload.

YewPar~\cite{Archibald2020} 
is a new combinatorial search framework designed to allow domain specialists to benefit from parallelism by reusing sophisticated parallel search patterns (\autoref{sec:YewPar}). As a performance baseline we develop  a direct exact sequential FSP solver in C++. 

The research contributions of this paper are as follows, and \textbf{the key results are summarised in \autoref{sec:Conclusions}}.

1. \textbf{We show that minimal effort is required to encode a typical OR problem  (FSP) in YewPar.} The direct and sequential YewPar solvers implement a standard FSP algorithm, and both require around 390 source lines of code (SLOC), much of which is shared. Although a simple Travelling Salesperson is reported in~\cite{Archibald18}, FSP is the first non-trivial OR application implemented using YewPar (\autoref{sec:implementation}). 

2.  \textbf{We show that YewPar makes it extremely easy to experiment with three sophisticated search parallelisations on shared and distributed memory architectures.} An alternate parallelisation simply entails a few lines of code that configure an appropriate YewPar skeleton, and the selection of suitable parameters. While \textit{StackStealing} \cite{Gmys2016} and \textit{DepthBounded} search coordinations \cite{Kouki2013} have been widely used, we report \emph{the first use of \textit{Budget} parallel search coordination for FSP}  (\autoref{sec:Parameter Tuning}). 

3. \textbf{We validate the solvers using 100+ standard instances; and baseline the sequential performance} against the recent LL solver \cite{Gmys2016} 
showing comparable performance.
YewPar is on average 5\% faster than LL, but 9\% slower than the direct FSP solver (\autoref{sec:validation}).

4.  \textbf{We systematically measure the performance of the three parallel FSP searches on 10 Taillard instances with up to 240 workers} on a Beowulf cluster. We show that YewPar increases the number of instances that can be practically solved, e.g. one instance runtime falls from 2.4 days to 16 minutes. Good absolute speedups (up to $297\times$) are achieved for large instances (sequential runtimes between 4 and 58 hours). Of the three parallelisations \textit{Budget} delivers the best performance, with mean speedup over 6 large Taillard instances of 246$\times$ 
(\autoref{sec:Comparing Parallel}). 

\section{Background}

Job-Shop scheduling optimises the scheduling of $n$
jobs onto $m$ machines while maintaining job precedence constraints. In Flowshop scheduling the precedence constraints are
identical for all jobs and the goal is to minimise the \emph{makespan}---the time to complete all jobs. There are a variety of solution
techniques~\cite{allahverdi_SchedulingSurvey:2016} and here we seek exact solutions that both
identify a
solution \emph{and} prove that no better solution exists.
One exact technique uses a branch-and-bound search of the tree of all possible solutions, with a bounding function that eliminates sub-trees that cannot contain an improved solution. Each search tree node represents a (partial) schedule, and the tree is never fully manifest in memory: it is \emph{generated} during the search.

Parallelising the search can reduce runtimes by concurrently exploring  sub-trees. While there is huge scope for parallelism, parallel tree search is far from trivial. We need to
determine \emph{how} and \emph{when} a tree should be split; subtree sizes vary hugely and work must be balanced between cores; global
knowledge, \eg new bounds, must be shared. The parallel coordination must
be added to the  search code, and is often so intrusive that only a single parallelisation is developed. Moreover parallelisation typically targets the less-challenging shared-memory architectures, but this limits scalability to around 100 cores.

\subsection{YewPar} 

YewPar is the \emph{``the first general purpose scalable parallel framework for exact combinatorial search''}~\cite{Archibald2020} and is designed to make it easier to
parallelise search applications.
\label{sec:YewPar}
YewPar offers reusable generic
parallel search patterns (algorithmic skeletons), that only require user parameterisation with sequential functions. It has been used to parallelise 7 different search applications using 3 different
parallelism techniques~\cite{Archibald2020}. 
Other general search frameworks include TaskWork~\cite{kehrer.ea_Taskwork:2019} and MaLLBa~\cite{alba.almeida.ea_MALLBAALibraryOfSkeletons:2002}, but these support only a single search coordination.

To separate search tree generation from parallelism (tree exploration) a
YewPar user specifies their search application as a \emph{Lazy Node Generator}
(LNG). A LNG specifies how, for a given node, a heuristically ordered list of child nodes is created. Lazy generation minimises memory overheads and avoids generating subtrees that are pruned.

\subsection{Search Coordinations}\label{sec:coordinations}
Search coordinations specify parallel work generation and distribution
strategies. YewPar currently supports four well-known search coordinations as follows. Detailed specifications of the YewPar search coordinations, including a parallel operational semantics for each, are available in~\cite{Archibald2020}.  

\textbf{Sequential} performs a depth-first search from the root of the search tree. It  executes in a single thread and is useful for performance baselining and debugging LNG implementations. 

\textbf{Depth-Bounded} creates a new task for any search tree node higher than a user-specified cut-off depth $d_{\mathit{cutoff}}$ with tasks shared  using locally-biased distributed random work-stealing.  The intuition is that tasks created near the root of the search tree will undertake significant searches, i.e. have long runtimes. Selecting a depth enables the user to control the number of tasks created.

\textbf{Budget} creates new tasks to be stolen every time a search task executes $budget$ back-tracks. The intuition is that search tasks that have a long runtime can usefully spawn other search tasks to help (the task has not instantly been pruned for example). Selecting a budget enables the user to control the size of the search tasks.

Budget is similar in spirit to restarting searches, but instead of starting the search afresh after a timeout (a budget), it creates new  tasks that continue the search from the (top-of the) current subtree.  Unlike a restart these new tasks do not necessarily execute immediately, as they may be queued waiting for a worker to become available.
Parallelising searches using restarts is an ongoing research focus, e.g.~\cite{archibald_sip}.

\textbf{Stack-Stealing} is a dynamic work generation approach where idle worker-threads request work directly from other workers (aka. work stealing). On receipt of a work-stealing request a worker may either respond that it has no work available, or send a node the search tree as close to the root as possible. That is, it also assumes that tasks created near the root of the search tree have long runtimes, and hence are worth communicating.

\subsection{Other Parallel FSP searches} 
Many parallel FSP searches are handcrafted, in contrast our impementation uses the general purpose YewPar framework.  A master-worker paradigm is common where tasks and new knowledge are stored on a \emph{centralised} master and distributed dynamically to workers, e.g.~\cite{Chakroun2016,Kouki2013}. To overcome the centralisation bottleneck   decentralised approaches use work-stealing, not unlike the StackStealing coordination, e.g.~\cite{Vu2012,Vu2016}. 

Some recent permutation FSP implementations have obtained excellent performance on GPUs, sometimes exploiting $10^{4}$ GPU cores e.g.~\cite{GymsGPU}. Some of these approaches exploit a specific \emph{Integer-Vector-Matrix} (IVM) data-structure to enable efficient parallelism both on CPUs and hybrid CPU/GPU architectures, e.g.~\cite{Gmys2016,Gmys2020}. Compared with YewPar, GPU and hybrid CPU/GPU programming requires significantly more developer effort.

Crucially previous parallelisations are similar to DepthBounded and StackStealing search coordinations, but \emph{we are not aware of any prior parallelisation using a Budget coordination}. 

\section{FSP Implementations}\label{sec:implementation}

We engineer two entirely standard FSP solvers for comparison purposes: a \emph{direct} sequential C++ implementation, and a YewPar lazy node generator that supports sequential, and three parallel searches: one for each coordination. Our implementations~\cite{zenodo} are  branch-and-bound and use the common approach of maintaining two partial schedules $\sigma_{1}$ (initial schedule) and $\sigma_2$ (final schedule)~\cite{Ladhari,Gmys2020}.
They are competitive rather than state-of-the-art FSP solvers.

Branching adds an unscheduled job (chosen in ascending numerical order) $j$ to either $\sigma_{1}$ or $\sigma_{2}$. The choice of $\sigma_{1}$ or $\sigma_{2}$ is based on the \emph{Alternate} rule~\cite{Gmys2020} that adds $j$ to $\sigma_{1}$ if the current tree depth is odd, else $\sigma_{2}$. We use the simple \emph{One Machine Bound} of Ignall and Schrage~\cite{schrage} that has low runtime at the cost of some tightness. 

Finally, we use the NEH approximation algorithm~\cite{NEH} to compute an initial upper bound before search.

\begin{cpp}[caption=YewPar StackStealing FSP Search,label={lst:Stack-stealing}]
sol = StackStealing<GenNode, Optimisation, BoundFunction<bound_func>, ObjectiveComparison<std::less<unsigned>>>::search(space, root, params);
\end{cpp}
% \begin{lstlisting}[caption=YewPar StackStealing FSP Search,label={lst:Stack-stealing},captionpos=t,float,abovecaptionskip=-\medskipamount]
% sol = StackStealing<GenNode, Optimisation,
%                     BoundFunction<bound_func>,
%                     ObjectiveComparison<std::less<unsigned>>>
%                     ::search(space, root, params);
% \end{lstlisting}

\begin{cpp}[caption= YewPar Budget FSP Search,label={lst:Budget}]
params.backtrackBudget = opts["backtrack-budget"].as<unsigned>();
sol = Budget<GenNode, Optimisation, BoundFunction<bound_func>, ObjectiveComparison<std::less<unsigned>>>::search(space, root, params);
\end{cpp}

% \begin{lstlisting}[caption= YewPar Budget FSP Search,label={lst:Budget},captionpos=t,float,abovecaptionskip=-\medskipamount]
% params.backtrackBudget = opts["backtrack-budget"].as<unsigned>();
% sol = Budget<GenNode, Optimisation,
%              BoundFunction<bound_func>,
%              ObjectiveComparison<std::less<unsigned>>>
%              ::search(space, root, params);
% \end{lstlisting}

\subsection{Comparing YewPar and Direct Solver Programs} \label{sec:Comparing}
The direct implementation requires some 370 source lines of code (SLOC) \cite{zenodo}, as counted with CLOC.
In comparison, the YewPar Sequential implementation requires 390 SLOC \cite{zenodo}. It is easy to refactor an existing search to a Lazy Node Generator as much of the code, \eg bounding functions is shared. Parallelising a YewPar search entails specifying reusable serialisation functions (30 SLOC). Thereafter specifying an alternate parallel search requires only a few lines of code, and  Listings~\ref{lst:Stack-stealing} and~\ref{lst:Budget} show the YewPar stack stealing and budget FSP parallelisations.

\section{Validation, Baselining and Parameter Tuning} \label{sec:validation}

\subsection{Experimental setup}\label{sec:Experimental setup}
Experiments are run on a Beowulf cluster consisting of 16 nodes (256 cores). Each node uses Ubuntu 14.04.3 LTS, has 2 Intel Xeon E5-2640v2 2GHz CPU (no hyper-threading), 64GB of RAM, and a 10Gb Ethernet Interconnect. For YewPar each node dedicates one core to management tasks meaning that each node has 15 worker threads (workers) doing sub-tree search.

Performance analysis of parallel searches is notoriously difficult due to the non-determinism caused by pruning, finding alternate valid solutions, and random work-stealing. These lead to performance
anomalies~\cite{debruin.kindervater.ea_AsyncrhoousParBnBandAnomalies:1995} that manifest as slowdowns or superlinear speedups.
We control for this by investigating multiple FSP instances, running each experiment multiple times, and reporting cumulative statistics.

\begin{table*}[]
  \centering\footnotesize
  \caption[]{\label{fig-eg} 10 Taillard validation examples, and baselining YewPar against the direct and LL solvers~\cite{Gmys2016}.}
 \begin{tabular}{@{}l r r r r r r r@{}}Instance & \makecell{Make-\\ span}  & \makecell{Direct solver \\ runtime(s) }   &  \makecell{YewPar \\ runtime(s) } & \makecell{YewPar:Direct \\ Slowdown} & \makecell{LL \\ runtime(s)} & \makecell{YewPar:LL \\ Slowdown} \\ \midrule
Ta21 & 2297 & 96590 & 104928 & 8.6\% & 24489 & 328.5\% \\
Ta22 & 2099 & 5039 & 5479 & 8.7\% & 11758 & -53.4\% \\
Ta23 & 2326 & 192641 & 208976 & 8.5\% & 79322 & 163.5\% \\
Ta24 & 2223 & 12732 & 13758 & 8.1\% & 19753 & -30.3\% \\
Ta25 & 2291 & 16425 & 17963 & 9.4\% & 25332 & -29.1\% \\
Ta26 & 2226 & 32744 & 35302 & 7.8\% & 34562 & 2.1\% \\
Ta27 & 2273 & 39013 & 42421 & 8.7\% & 28295 & 49.9\% \\
Ta28 & 2200 & 1022 & 1119 & 9.5\% & 4569 & -75.5\% \\
Ta29 & 2237 & 1546 & 1688 & 9.2\% & 3674 & -54.1\% \\
Ta30 & 2178 & 1139 & 1237 & 8.6\% & 898 & 37.8\% \\\bottomrule
\makecell[l]{Geometric \\ mean} & & & & 8.7\% & & -4.72\% \\

\end{tabular}
\label{table:nodes decomposed}

\end{table*}

\subsection{Validation} \label{sec:Validation.} The "correctness" of the direct and YewPar FSP solvers is validated by comparing their makespans with published results for 107 well-known instances including the Carlier instances~\cite{Carlier}, the first thirty Taillard instances~\cite{Taillard}, and VRF instances \cite{VRF} up to 20 jobs, 15 machines, excluding one instance with runtime over 12 hours \cite{zenodo}. 
\cref{table:nodes decomposed} illustrates some Taillard instances (21-30),  that have significant runtime and published performance results~\cite{Gmys2016}. 
For all instances the direct and YewPar solvers visits an equal number of search tree nodes, showing that the search algorithms are nearly identical.

\subsection{Sequential Performance Baselining} 
\label{sec:Sequential Performance Base-lining}

Columns 3, 4 and 5 of \autoref{table:nodes decomposed} report the direct and YewPar solver runtimes, and the YewPar slowdown. As YewPar provides reusable search encodings we expect it to have some overheads compared with a search-specific solver like the direct solver~\cite{Archibald2020}. 
For these 10 instances the observed slowdown varies from 7.8\% to 9.5\%  with a geometric mean of 8.7\%. 

To relate YewPar performance with an existing algorithm Columns 6 and 7 of \autoref{table:nodes decomposed} compare the runtimes of sequential YewPar with a recent linked-list (LL) solver~\cite{Gmys2016} for Taillard instances 21-30~\cite{Taillard}. The hardware/OS platforms are similar, specifically the core has the same 2GHz clock frequency (on a pair of 8-core Sandy Bridge E5-2650 processors), uses 32 GB of memory, and CentOS 6.5 Linux~\cite{Gmys2016}.

\begin{figure}[]
  \centering
  \includegraphics[width=\linewidth]{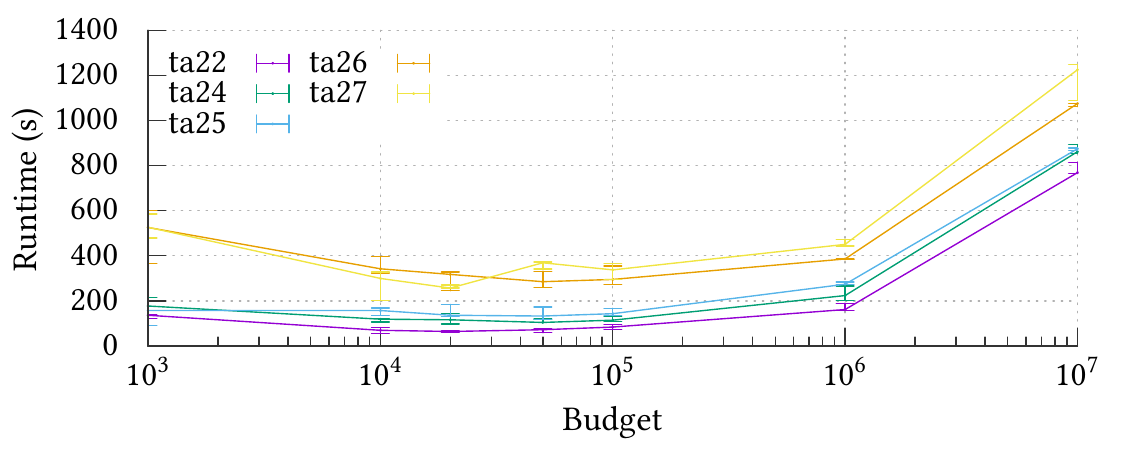}
 \caption{Budget parameter sweep: runtime on 120 workers}
 \label{fig:Budget parameter}
\end{figure}

As expected the comparative runtimes vary greatly: exactly half of the YewPar runtimes are shorter than LL. The differences are almost certainly due to the solvers using different bounding operators: YewPar uses a computationally cheap \textit{One Machine Bound} (\autoref{sec:implementation}) while LL uses a more expensive, but tighter \textit{Two Machine Bound}~\cite{Gmys2016}. A One Machine Bound may extend runtime if it fails to prune significant subtrees, but being computationally inexpensive may reduce runtimes in searches where tight bounds are less important. 

As before we expect that the generic YewPar solver will incur overheads compared with a search-specific solver like LL. YewPar slowdowns vary from -75.5\% to 328.5\%. The geometric mean slowdown for these search instances is -4.72\%, and we conclude that the sequential performance of YewPar solver is comparable to state-of-the-art solvers.

\subsection{Parallel Search Parameter Tuning} \label{sec:Parameter Tuning}
Two YewPar parallel search coordinations require the selection of suitable parameter values, specifically the depth at which to spawn tasks in a \textit{DepthBounded} search, and the number of backtracks that a search task should perform  in a \textit{Budget} search before generating new search tasks. We determine suitable values from a parameter sweep using Taillard instances with sequential runtimes between 1 and 12 hours. To account for random work stealing the instances are measured three times on 120 workers, and we plot lower, median, and upper values.

The \textit{budget} parameter sweep initially considered exponentially increasing values between $10^3$ and $10^7$, as shown in  \autoref{fig:Budget parameter}. Runtimes fall between budgets $10^3$ - $10^4$, and increase  between $10^5$ - $10^7$. Additional measurements at  $2 \times 10^4$ and $5 \times 10^4$ reveal small differences, and we select a budget of  $5 \times 10^4$ for the evaluation in \autoref{sec:Comparing Parallel}.  The sweep for the \textit{depth} is similar, and identifies depth 5 as providing the best performance for these searches.

\section{Comparing Three Parallel FSP Searches} 
\label{sec:Comparing Parallel}

\begin{figure}
  \centering
\includegraphics[width=\linewidth]{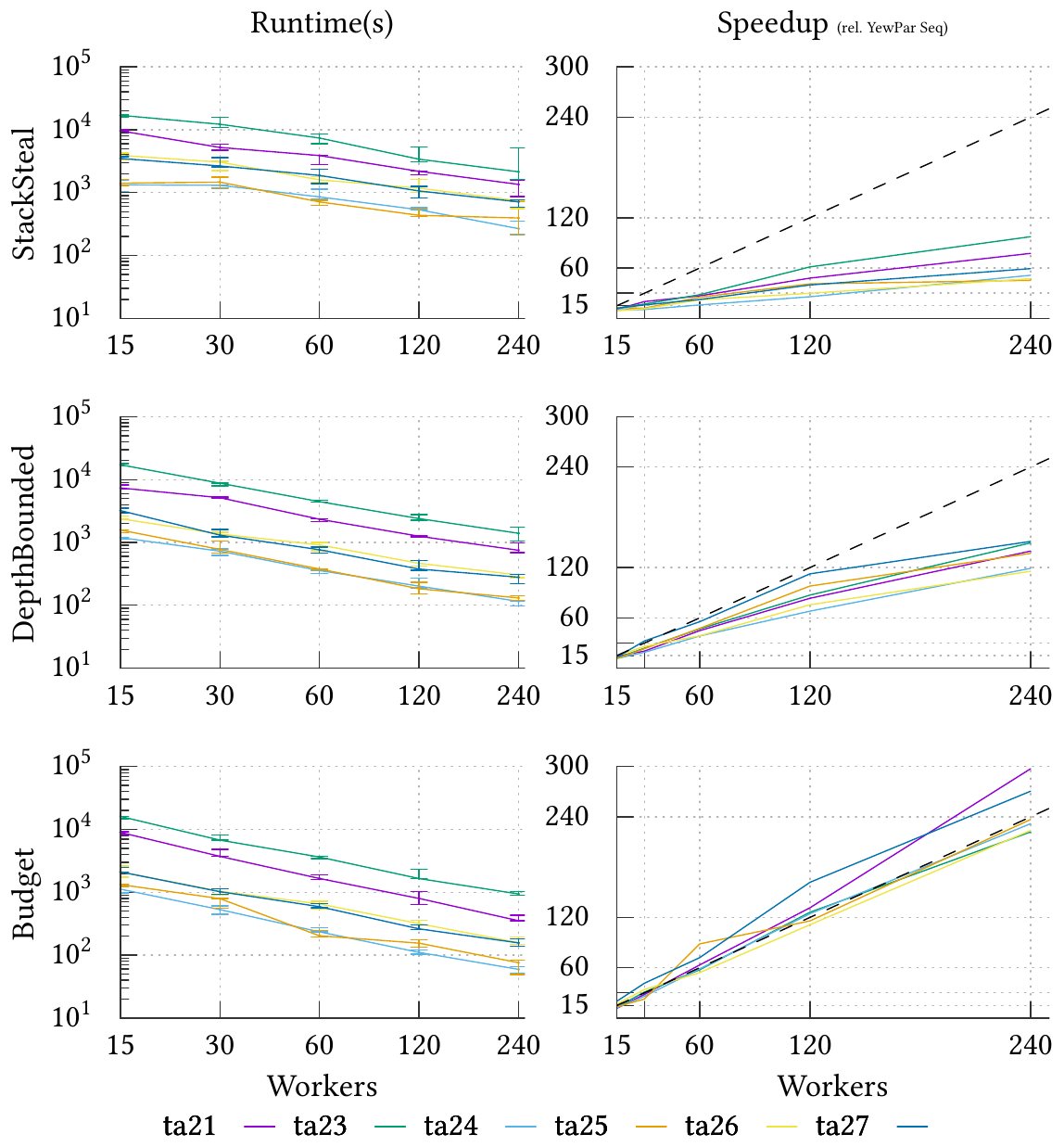}

  \caption{Runtime and Speedups for 3 parallel flowshop searches: StackStealing, DepthBounded and Budget.}
  \label{fig:results}
\end{figure}

A key benefit of the YewPar framework is that experimenting with alternative parallel coordinations is easy, requiring limited code changes (\cref{sec:Comparing}). This is essential as previous work~\cite{Archibald18} shows  that there is no best coordination for different search applications.

This section compares the performance of \textit{StackStealing}, \textit{DepthBounded} and \textit{Budget} YewPar FSP searches. 
We again select the Ta21-30 instances, and group them based on the sequential YewPar runtimes: \textit{small} instances complete within 3.5 hours,  and  \textit{large} within 60 hours on our cluster. As is typical for parallel workloads, the \emph{small} instances have least scope to scale as runtimes are already low on small numbers of cores, so we initially focus on the large Taillard instances, i.e.\  Ta21--27 excluding Ta22. 
The performance of the three parallel search coordinations is shown in \cref{fig:results}.
Speedups are relative to sequential YewPar runtime, and we plot ideal speedup as a dashed line. Runtimes and speedups are shown on log-log plots, with error bars representing the min/max measured runtimes.

\subsection{StackStealing Evaluation (top plots)} \label{sec:Stack-stealing Evaluation}
While StackStealing requires no parameter tuning, it does not scale particularly well and quickly diverges from linear speedup.
It is likely that many tasks are pruned early and the hardest subtrees remain on a single cluster node (effectively giving 15 workers). As steals are random it seems that remote workers don't locate hard tasks.

The range between minimum and maximum runtimes is often large (note log scale), as expected due to random work-stealing. Despite these issues, without exploiting a large cluster and without  parameter tuning, runtimes are significantly reduced. For example the median runtime of Ta23 falls from 4.7 hours on 15 workers (58 hours on a single worker) to less than an hour on 240 workers.

\subsection{DepthBounded Evaluation (middle plots)} \label{sec:Depth-bounded Evaluation} The DepthBounded searches use the cutoff depth  $d_{\mathit{cutoff}} = 5$ determined to be effective (\autoref{sec:Parameter Tuning}).
DepthBounded search achieves good speedups for all instances, with no apparent limit to the scaling. As DepthBounded spawns new tasks only as a task at depth $< d_{\mathit{cutoff}}$ is \emph{executed}, a single steal may generate multiple tasks improving load balance, and hence speedups. This is especially true for the relatively low $d_{\mathit{cutoff}}$ values used here. This is also likely why the runtime ranges are relatively small compared to StackStealing despite random work distribution.

\subsection{Budget Evaluation (bottom plots)} \label{sec:Budget Evaluation} The searches use a budget of $5\times10^{4}$ backtracks that was determined to be effective (\autoref{sec:Parameter Tuning}).
Speedups are excellent: near linear  in all cases, with some super-linear speedups due to knowledge transfer and speculation.
We believe that Budget performs so well because it is able to generate only tasks that contain large amounts of work. Clearly this task size heuristic is more effective than the heuristics used in the other coordinations, e.g.~depth. 
As for DepthBounded the range of runtimes is small despite random work distribution.

\begin{table*}[]
  \centering\footnotesize
\caption{Comparing parallel search coordinations on \textbf{Small} FSP instances with 15 or 240 workers. Smaller speedups than for larger instances, and Budget outperforms DepthBounded and StackStealing (runtimes in seconds).}
 \begin{tabular}{@{}lrrrrrrrrrr@{}} & Sequential & \multicolumn{3}{c}{StackStealing} & \multicolumn{3}{c}{DepthBounded} & \multicolumn{3}{c}{Budget} \\
& runtime & \multicolumn{2}{c}{runtime} & spdup & \multicolumn{2}{c}{runtime} & spdup
 & \multicolumn{2}{c}{runtime} & spdup \\
 \cmidrule(r){2-2}
 \cmidrule(r){3-5}
 \cmidrule(r){6-8}
  \cmidrule(r){9-11}

 Instance & 1 & 15 & 240 & 240 & 15 & 240 & 240 & 15 & 240 & 240 \\ \midrule
 Ta22 & 5479 &  642 & 133  & 41 & 590 &  68 & 81 & 463 & 55 & 100 \\
 Ta28 & 1119 &  102 & 31   & 36 & 252 &  19 & 59 &  80 & 12 & 93  \\
 Ta29 & 1688 &  179 & 17   & 99 & 330 &  32 & 53 & 112 & 19 & 89  \\
 Ta30 & 1237 &  142 & 20   & 62 & 281 &  22 & 56 &  95 & 19 & 65  \\
 \midrule

 \makecell[l]{Geom. \\ mean}  & & &  & 55 & & & 61  & & & 86 \\

\end{tabular}
\label{table:small-instances}
\end{table*}

\subsection{Small Search Instances} \label{sec:SmallerInstances}

While large search instances provide the best parallel performance, YewPar can also be applied to smaller search instances. \cref{table:small-instances} summarises the performance of the three parallel search coordinations for the four small Taillard instances, i.e.\ with sequential runtimes less that 3.5 hours. It shows runtimes in seconds and speedups relative to sequential YewPar runtime with 15 and 240 workers. 15 is the number of workers on a single shared-memory cluster node, and 240 workers is the maximum number of workers on the 16-node cluster. 

Runtimes on small numbers of cores are already short, e.g.\ around 10 minutes on 15 cores, and hence the speedups are lower than for larger instances. Nevertheless all three search coordinations achieve mean speedups in excess of 50 for these 4 instances.

\begin{table*}[]
  \centering\footnotesize
\caption{Comparing parallel search coordinations on \textbf{Large} FSP instances with 15 or 240 workers. Budget outperforms DepthBounded and StackStealing (runtimes in seconds).}
 \begin{tabular}{@{}lrrrrrrrrrr@{}} & Sequential & \multicolumn{3}{c}{StackStealing} & \multicolumn{3}{c}{DepthBounded} & \multicolumn{3}{c}{Budget} \\
& runtime & \multicolumn{2}{c}{runtime} & spdup & \multicolumn{2}{c}{runtime} & spdup
 & \multicolumn{2}{c}{runtime} & spdup \\
 \cmidrule(r){2-2}
 \cmidrule(r){3-5}
 \cmidrule(r){6-8}
  \cmidrule(r){9-11}

 Instance & 1 & 15 & 240 & 240 & 15 & 240 & 240 & 15 & 240 & 240 \\ \midrule

 Ta21 & 104928 & 9620 & 1354 & 78 & 7359 & 751 & 140 & 8988 & 353 & 297 \\
 Ta23 & 208976 & 17009 & 2145 & 97 & 17251 & 1397 & 150 & 16141 & 941 & 222 \\
 Ta24 & 13758 & 1331 & 268 & 51 & 1196 & 116 & 119 & 1122 & 59 & 233 \\
 Ta25 & 17963 & 1415 & 395 & 46 & 1579 & 131 & 137 & 1325 & 76 & 236  \\
 Ta26 & 35302 & 3914 & 745 & 47 & 2393 & 305 & 116 & 2117 & 158 & 223 \\
 Ta27 & 42421 & 3501 & 715 & 59 & 3219 & 280 & 152 & 2112 & 157 & 270 \\
 \midrule

 \makecell[l]{Geom. \\ mean}  & & &  & 61 & & & 135  & & & 246 \\

\end{tabular}
\label{table:coord-comparison}
\end{table*}

\subsection{Parallel Coordination Comparison} \label{sec:Parallel Coordination Comparison} \cref{table:coord-comparison} summarises the performance of the three parallel search coordinations for the 6 large FSP instances. We compare the performance of the search coordinations considering both large and small FSP instances (\cref{table:small-instances}). 

For both large and small instances Budget outperforms the other coordinations, with significantly reduced runtimes, and the highest speedups. For example the geometric mean speedup for the large instances is more than 4 times that of StackStealing: 245$\times$ compared with 61$\times$.  For the four smaller instances it provides mean speedups over 80$\times$. Indeed the speedups for 2 of the 10 instances are super-linear with a maximum speedup of 297$\times$. 

\textit{DepthBounded} is next best, for the larger instances it consistently provides speedups over 115$\times$ and a mean speedup of 135$\times$, and for the  smaller instances speedups over 60$\times$.  \textit{StackStealing} provides the worst performance with not a single instance reaching 100$\times$ speedup, and the mean for the 6 larger instances being 61$\times$.

\section{Conclusions} \label{sec:Conclusions}

We report the first YewPar encoding of a non-trivial OR problem: a standard FlowShop Scheduling solver.
The encoding is low effort: around 390 source lines of code (SLOC), compared with 370 for a direct FSP implementation  (\autoref{sec:Comparing}). Alternate parallelisations simply entail (1) some reusable serialisation (30 SLOC), and (2) configuring an appropriate skeleton from the YewPar library: a few lines of code and the selection of suitable parameters, e.g.\ compare Listings~\ref{lst:Stack-stealing} and~\ref{lst:Budget}.
Moreover the \textit{Budget} parallel search coordination is a first for FSP.  

Our solvers are validated with 107 standard (Carlier, Taillard, and VRF) FSP instances. We baseline the sequential performance of YewPar against the LL state-of-the-art solver \cite{Gmys2016} on a similar hardware/OS platform. There is considerable variation in the runtimes of the Ta21 - Ta30 searches (Columns 4\&6 of \autoref{table:nodes decomposed}) as the solvers use different bounding operators.
The performance of LL and YewPar is comparable for the searches: YewPar is faster in 5 out of 10 instances, and on average 5\% faster. 
There is, however, a performance penalty for YewPar's generality: a mean 8.7\% slowdown compared with the direct solver for the same 10 Taillard instances (Columns 3,4\&5 of \autoref{table:nodes decomposed}).

Systematic performance measurements of the three parallel FSP searches on a Beowulf cluster with 240 workers shows that YewPar increases the number of instances that can  practically be solved, e.g. the runtime of Ta21 falls from 2.4 days to 16 minutes. YewPar also provides useful speedups for small search instances, e.g. mean speedups of over 50$\times$ in \cref{table:small-instances}.
Of the three search coordinations, the novel \textit{Budget} coordination performs best on both large and small instances.
For example on the six larger Taillard instances it consistently provides absolute speedups of over 220$\times$ and a geometric mean speedup of 246$\times$.  
\textit{DepthBounded} is next best, for the larger instances it consistently provides speedups over 115$\times$ and a mean speedup of 135$\times$.
Although it requires no prior parameter tuning,  \textit{StackStealing} provides the worst performance with not a single instance reaching 100$\times$ speedup, and the mean for the 6 larger instances being 61$\times$ (Tables~\ref{table:small-instances} and~\ref{table:coord-comparison} in \autoref{sec:Parallel Coordination Comparison}).

\textbf{Future work} may compare the parallel performance implications of alternate bounding functions, e.g. the \textit{Two Machine Bound} \cite{Gmys2016}, and this can be done without any parallelism code changes. 
There are many very large FSP instances that require weeks to compute sequentially, and it would be interesting to explore whether these representative OR problems could be practically solved by deploying YewPar on a mid-scale HPC (around 10K cores).

\bibliography{bibliography}

\end{document}